\documentclass[aps,prx,twocolumn]{revtex4-2}
\usepackage{amsmath}
\usepackage{amssymb}
\usepackage{amsfonts}
\usepackage{graphicx}
\usepackage{dcolumn}
\usepackage{bm}
\usepackage{sidecap}
\usepackage{subfloat}
\usepackage{parskip}
\usepackage{grffile}
\usepackage{color}
\usepackage{hyperref}
\usepackage{hhline}
\usepackage{mathtools}
\usepackage{graphics}
\usepackage{multirow}
\usepackage{verbatim}
\usepackage{longtable}
\usepackage{rotating}
\usepackage{setspace}
\usepackage{epsfig}
\usepackage{subfigure}
\usepackage{epstopdf}
\usepackage{gensymb}
\usepackage[normalem]{ulem}
\usepackage{tikz}
\usepackage{physics}
\graphicspath{{./}{ER/main/}}

\begin{document}


\title{Stochastic Multipath Routing for High-Throughput Entanglement Distribution in Quantum Repeater Networks}
\author{Ankit Mishra} \email{ankitphy0592@gmail.com}
\author{Kang Hao Cheong} \email{kanghao.cheong@ntu.edu.sg (Corresponding Author)}
\affiliation{School of Physical and Mathematical Sciences, Nanyang Technological University, 21 Nanyang Link, S637371, Singapore}

\begin{abstract}
Quantum repeater networks distribute entanglement over lossy links while many users share a limited pool of entangled pairs. Most existing routing schemes either always use a single best path or rely on global optimizations that are hard to run in real time. Here we propose and analyze a simple alternative: a stochastic multipath rule in which each entanglement request is sent at random along one of several edge-disjoint repeater paths, with a single parameter that controls the bias between shorter and longer routes. Using a distance-dependent lossy network model with finite per-link capacities and probabilistic entanglement swapping, we develop an analytic description of the resulting end-to-end entanglement rate as a function of this bias and validate it with large-scale numerical simulations. We find that an intermediate bias consistently outperforms both deterministic extremes across distances, traffic patterns, attenuation, swapping noise, and congestion, bringing the rate close to simple capacity upper bounds and making link usage more even across networks. These results identify stochastic multipath routing as a lightweight classical control strategy for boosting performance and scalability in near-term quantum repeater networks.
\end{abstract}

\maketitle

 Quantum networks distribute entanglement between distant nodes, crucial for quantum communication, computation, and sensing \cite{azuma2023quantum, wehner2018quantum}. Network nodes act as quantum processors and repeaters, generating, storing, and processing quantum information via optical links \cite{shi2020concurrent,azuma2023quantum}. Quantum communication requires establishing entanglement between source (Alice) and destination (Bob), enabling quantum information transmission in the form of qubits. However, quantum communication is highly susceptible to losses and noise, limiting direct communication distances and rates \cite{pirandola2009direct,takeoka2014fundamental,bapat2023advantages}. In particular, Ref.~\cite{pirandola2017fundamental} establishes a fundamental upper bound on quantum communication rates, $C(\eta) = -\log_{2}(1-\eta)$, saturating near $1.44 \eta$ ebits per channel use for small transmissivity $\eta <<1$. Quantum repeaters are thus introduced, similar to classical link-layer repeaters, convert exponential entanglement generation costs to polynomial, enabling scalable quantum networks \cite{briegel1998quantum,duan2001long,childress2005fault}. Quantum repeaters serve as intermediate quantum stations that extend entanglement range via entanglement swapping, typically realized through two-qubit Bell-state measurements (BSMs) \cite{yuan2008experimental,bernien2013heralded,chou2005measurement}. By stitching short-distance entangled links into long-range connections, BSMs enable scalable entanglement distribution. However, BSMs remain inherently probabilistic due to photon loss, hardware imperfections, and limited detection efficiency—constrains fidelity and success rates. Consequently, building large-scale quantum networks to support dynamic Source-Destination pair requests demands efficient entanglement path selection and judicious use of constrained resources, such as limited entangled pairs per link.

Fundamental end-to-end capacity limits for repeater chains and quantum networks under both single-path and multi-path routing were established in Ref.~\cite{pirandola2019end}, which showed in particular that multipath use can strictly outperform single-path routing in lossy networks. Building on these capacity-level insights, subsequent works have developed practical routing strategies for entanglement distribution under edge-disjoint path selection, fidelity constraints, entanglement failures, and related network-level resource constraints\cite{pant2019routing,li2021effective,zeng2023entanglement,li2022fidelity,gyongyosi2017entanglement,harney2025practical}.
These multipath schemes are typically realized through centralized controllers and heavy global optimization (e.g., flow formulations or integer programs) that select one globally best route or attempt to activate all available paths subject to fidelity or rate constraints.  In this work, we focus on how traffic is split between \emph{short} and \emph{long} edge-disjoint routes in a realistic noisy repeater network, and show that a very simple randomized tournament over all disjoint paths exhibits a robust interior bias $0<\gamma^\star<1$ that mixes shorter and longer paths and consistently outperforms both extremes (favoring only the shortest paths, or favoring only the longer paths) for single and multiple concurrent source–destination pairs. At the same time, we derive and validate a compact analytic expression for the expected swap-weighted throughput $T_r(\gamma)$, so performance in this near-term regime can be predicted and tuned analytically rather than only by simulation. Put differently, relative to existing multipath routing schemes, our results add a new physics-based design principle: Parrondo-type mixing between short and long routes is generically near-optimal in noisy quantum repeater networks, and it can be implemented with minimal classical control overhead. This routing behaviour provides a concrete network-level realization of Parrondo's paradox, where alternating between individually lossy strategies yields a net performance gain \cite{harmer2002parrondo,mishra2024parrondo,mishra2024efficient,cheong_1,cheong_2,cheong_3}.

We model the quantum network as a random geometric graph (RGG) $G(V, E)$, where $N = |V|$ nodes, each equipped with a quantum repeater, are uniformly distributed over the unit square $[0,1]^2$. An undirected edge $(i,j) \in E$ exists between any two nodes $i$ and $j$ if their Euclidean separation $L_{ij}$ satisfies $L_{ij} \leq r$, where $r$ is the connection radius. The average degree of such a graph scales as $\langle k \rangle \approx N \pi r^2$, and the graph transitions from disconnected to connected with high probability when $r \gtrsim \sqrt{\frac{\log N}{\pi N}}$, ensuring global connectivity in the large-$N$ limit. Each edge $(i,j)$ represents a lossy optical quantum channel characterized by its physical length $L_{ij}$ and an associated transmissivity given by

\begin{equation}
    p_{\mathrm{link}}^{(ij)} = \exp(-\alpha L_{ij}),
\end{equation}

where $\alpha$ is the photon attenuation coefficient, typically expressed in inverse length units and related to the material and wavelength properties of the fiber. The exponential decay models distance-dependent photon loss and scattering in optical fibers, reducing entanglement success with increasing separation. Furthermore, each edge $(i,j)$ is equipped with a buffer capable of storing up to $C_0$ entangled pairs per time window. During each window, $C_0$ entanglement generation attempts are made independently across the link, each with success probability $p_{\mathrm{link}}^{(ij)}$. Consequently, the number of successfully established entangled pairs across edge $(i,j)$ is modeled as a binomial random variable,
\begin{equation}
    C_{ij} \sim \mathrm{Binomial}(C_0,\, p_{\mathrm{link}}^{(ij)}).
\end{equation}

We assume that all successfully generated entangled pairs possess high fidelity and are immediately usable for quantum communication via entanglement swapping, without requiring additional purification or error correction.\\

At each time step $t$, a single source--destination (S--D) pair is selected uniformly at random, and a fixed number of entanglement routing requests $f_r$ are generated. Let $n_t$ denote the number of edge--disjoint paths for the selected S--D pair in the post-entanglement feasible network, i.e., paths for which every edge $(i,j)$ satisfies $C_{ij}\geq 1$. Denote these feasible paths by $P_1,\dots,P_{n_t}$, ordered by nondecreasing hop count. Each request is then routed using a probabilistic tournament-style path-selection rule defined by the ordered recursive bisection induced by this ordering: for any contiguous block of $m$ paths, the left subset contains the first $\lceil m/2\rceil$ paths and the right subset contains the remaining $\lfloor m/2\rfloor$ paths. At each branching step, the left subset is selected with probability $\gamma$ and the right subset with probability $1-\gamma$, and the recursion continues until a single path is selected. If the selected path has sufficient residual capacity, i.e., at least one available entangled pair on each of its links, the request is accepted and consumes one entangled pair per edge along the path; otherwise, it is dropped. The total throughput in a given time window is defined as the sum of the swap-success probabilities of all accepted requests, and is given by
\begin{equation}
T_r=\sum_{s=1}^{f_r} w_{i_s}\,\mathbf{1}\{\text{path } i_s \text{ accepted at } s\},
\end{equation}
where $w_{i_s}=p_{\mathrm{swap}}^{\,h_{i_s}-1}$ is the success weight of the selected path $i_s$, $h_{i_s}$ is the hop count of path $i_s$, $p_{\mathrm{swap}}$ denotes the success probability of a Bell-state measurement (BSM) during entanglement swapping, and the indicator function equals one if the request is accepted in that window. 

We now derive an analytical expression for the expected throughput $T_r$. Conditioning on a fixed path count $n_t=n$, we first derive the conditional expectation $\mathbb{E}[T_r(\gamma)\mid n_t=n]$ and then obtain the unconditional expectation $\mathbb{E}[T_r(\gamma)]$ by the law of total expectation. For fixed $n$, let $P_1,\dots,P_n$ denote the feasible paths ordered by nondecreasing hop count, as defined above. If the leaf associated with path $P_i$ in the corresponding ordered recursive bisection tree lies at depth $k_i$ and its root-to-leaf trace contains $m_i$ left turns, then the per-request selection probability is
\begin{equation}\label{eq:pi_gamma_supp}
p_i(\gamma)=\gamma^{\,m_i}\,(1-\gamma)^{\,k_i-m_i},
\qquad
\sum_{i=1}^{n} p_i(\gamma)=1.
\end{equation}
With \(f_r\) independent requests per window the selection count for path \(i\) satisfies
\begin{equation}\label{eq:Ni_supp}
N_i \sim \mathrm{Binomial}\bigl(f_r,\;p_i(\gamma)\bigr).
\end{equation}

\begin{figure}[!htbp]
	\centering
	\includegraphics[width= 0.52\textwidth ]{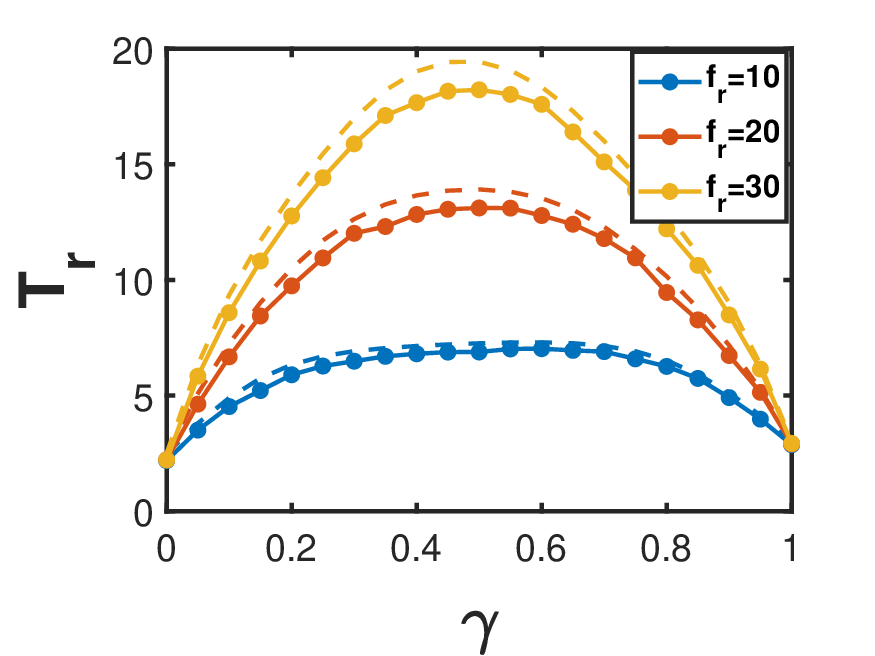} 
	\caption{Throughput $T_r$ versus tournament bias $\gamma$ for several request loads $f_r$. Solid circles show simulations (averaged over $T=1000$ windows); dashed curves show the analytical prediction. The other parameters are $r=0.105$, $C_0=5$, $p_{\mathrm{swap}}=0.95$, $\alpha =1$, $N=500$.}
	\label{Tr_sn_sim}
\end{figure}

Since each accepted routing of path \(i\) consumes one entangled pair from every hop on that path, the vector of hop-capacities \(\{C_{i,j}\}_{j=1}^{h_i}\) is reduced by one component-wise upon every acceptance. Consequently path \(i\) remains available only until the first hop along the path is exhausted; we therefore define the 
\emph{effective capacity}
\[
C_i^{\mathrm{eff}} =\; \min_{1\le j\le h_i} C_{i,j},
\]
which equals the maximum number of times path \(i\) can be accepted in the current time window. Thus, the expected throughput conditioned on $n$ can be calculated as
\begin{equation}\label{eq:ETr_sum}
\mathbb{E}[T_r\mid n_t = n] \;=\; \sum_{i=1}^n w_i\,\mathbb{E}\bigl[\min(N_i,C_i^{\mathrm{eff}})\bigr].
\end{equation}

Condition on \(C_i^{\mathrm{eff}}\) and use independence of selection counts and capacities:
\begin{equation}\label{eq:cond}
\mathbb{E}\bigl[\min(N_i,C_i^{\mathrm{eff}})\bigr]
=\sum_{c=0}^{C_0} \mathbb{E}\bigl[\min(N_i,c)\bigr]\;\mathbb{P}(C_i^{\mathrm{eff}}=c).
\end{equation}

To evaluate the right-hand side of Eq.\,\eqref{eq:cond}, introduce the tail-sum notation (valid for any integer \(c\ge0\))
\begin{equation}\label{eq:Sc_def}
S\!\bigl(f_r,p_i(\gamma);c\bigr)
=\mathbb{E}\bigl[\min(N_i,c)\bigr]
=\sum_{\ell=1}^{c}\mathbb{P}\bigl(N_i\ge \ell\bigr),
\end{equation}

\begin{figure}[!htbp]
	\centering
	\includegraphics[width= 0.52\textwidth ]{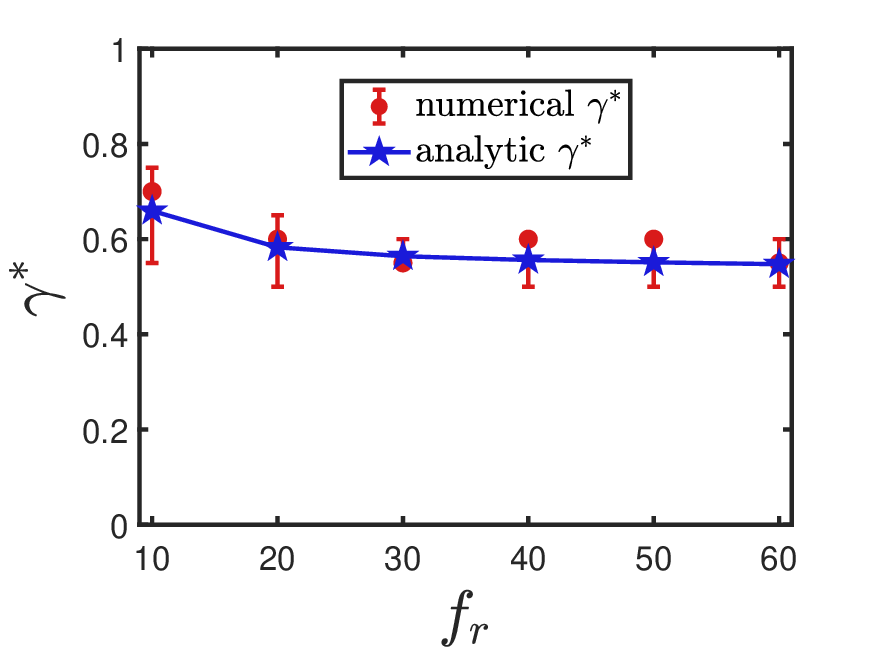} 
	\caption{Plot of the optimal tournament bias \(\gamma^{*}\) as a function of request loads $f_r$. The other parameters are the same as in Fig.~\ref{Tr_sn_sim}.}
	\label{gam_optimal}
\end{figure}

Next, each hop-capacity satisfies \(C_{i,j}\sim\mathrm{Binomial}(C_0,p_{\mathrm{link}}^{(ij)})\) and thus for a single-hop tail, we obtain
\begin{equation}\label{eq:Qij_repeat}
Q_{i,j}(c)=\mathbb{P}(C_{i,j}\ge c)
=\sum_{t=c}^{C_0}\binom{C_0}{t}\,(p_{\mathrm{link}}^{(ij)})^t(1-p_{\mathrm{link}}^{(ij)})^{C_0-t},
\end{equation}
with the convention \(Q_{i,j}(0)=1\).  Since the \(C_{i,j}\) are independent,
the mass function of \(C_i^{\mathrm{eff}}\) is thus the forward difference of these tails:
\begin{equation}\label{eq:Ceff_mass_repeat}
\mathbb{P}\bigl(C_i^{\mathrm{eff}}=c\bigr)
=\prod_{j=1}^{h_i} Q_{i,j}(c)\;-\;\prod_{j=1}^{h_i} Q_{i,j}(c+1),
\end{equation}
where \( \prod_{j=1}^{h_i} Q_{i,j}(C_0+1)=0\) by convention.

Substitute (\ref{eq:Sc_def}) and (\ref{eq:Ceff_mass_repeat}) into (\ref{eq:cond}) and then into (\ref{eq:ETr_sum}).  After reordering finite sums we obtain the closed-form expression for conditional expected throughput
\begin{equation}\label{exp_thr}
    \mathbb{E}[T_r(\gamma)\mid n_t = n]
=\sum_{i=1}^n w_i \sum_{c=0}^{C_0} S\bigl(f_r,p_i(\gamma);c\bigr)\, \mathbb{P}\bigl(C_i^{\mathrm{eff}}=c\bigr).
\end{equation}

Finally, by the law of total expectation, averaging over the random tournament size \(n_t\) yields the unconditional expected throughput
\begin{equation}\label{eq:total_expectation_final}
\mathbb{E}[T_r(\gamma)]
\;=\; \sum_{n\ge 2}\Pi_n\,\mathbb{E}\!\left[T_r(\gamma)\,\middle|\, n_t=n\right],
\qquad
\Pi_n=\mathbb{P}(n_t=n).
\end{equation}
It is also worth mentioning here that if the tournament is balanced so that all leaves have the same depth \(k_i=k^\star\), then for \(\gamma=\tfrac12\) one has \(p_i(\tfrac12)=2^{-k^{*}}=1/n\) and \eqref{exp_thr} simplifies with a common \(S(f_r,1/n;c)\). For arbitrary (unbalanced) trees, \(p_i(\tfrac12)=2^{-k_i}\) and the general form \eqref{exp_thr} applies.

The outcome of the Parrondo–style mixing protocol is shown in Fig.~\ref{Tr_sn_sim} where the swap-weighted throughput \(T_r\) is plotted versus the tournament bias \(\gamma\) for several request loads \(f_r\). Stochastic mixing between the two deterministic limits; always selecting the left (shorter-path) or right (longer-path) branch, i.e., \(\gamma=1\) and \(\gamma=0\) produces a pronounced interior optimum that surpasses either extreme. As \(f_r\) increases, the peak near \(\gamma \approx \tfrac{1}{2}\) becomes more prominent, reflecting that distributing traffic across the  edge-disjoint routes postpones saturation of the path bottlenecks \(C_i^{\mathrm{eff}}\) whereas the extreme policies concentrate load and exhaust those bottlenecks earlier. In the simulations, each time window uses all edge-disjoint paths for entanglement routing, and the throughput is averaged over windows. The analytical curve is evaluated from Eq.~(\ref{eq:total_expectation_final}) with a geometry–averaged hop success \(p=\langle e^{-\alpha L_{ij}}\rangle_{(i,j)\in E}\) and hop counts \(h_i\simeq h_1+c\,i^{\beta}\) fitted on the post-entanglement geometry over a large ensemble. To obtain the optimal tournament bias we differentiate Eq.~\ref{eq:total_expectation_final} with respect to \(\gamma\) and get

\begin{figure}[!htbp]
\centering
\includegraphics[width=1.05\columnwidth]{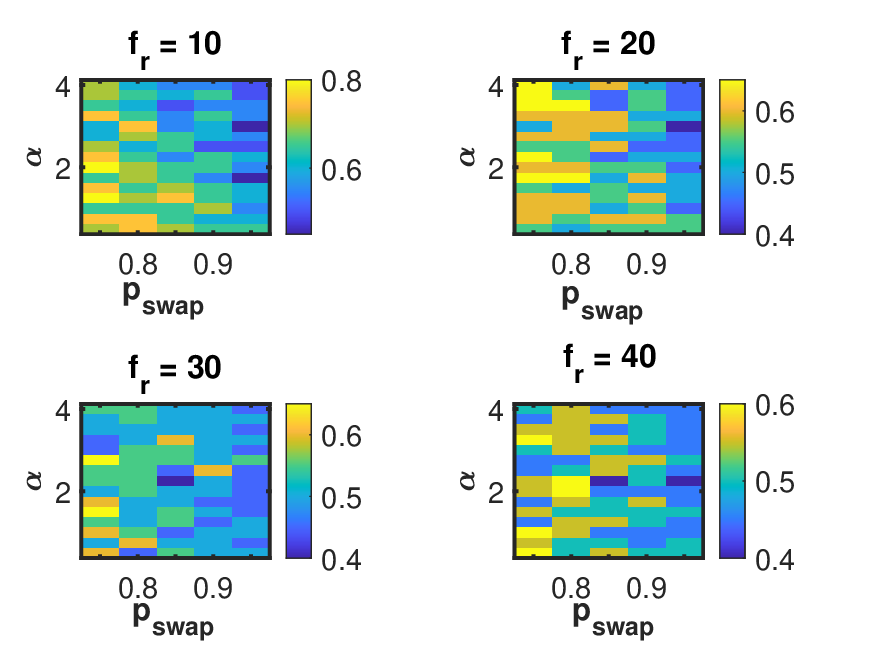}
\caption{Heat map of the optimal tournament bias \(\gamma^{*}\) as a function of swap success \(p_{\mathrm{swap}}\) and attenuation \(\alpha\) for various values of \(f_r\) while other parameters are the same as in Fig.~\ref{Tr_sn_sim}.}
\label{fig:gam_opt_alph_pswp}
\end{figure}

\begin{figure}[!htbp]
\centering
\includegraphics[width=0.95\columnwidth]{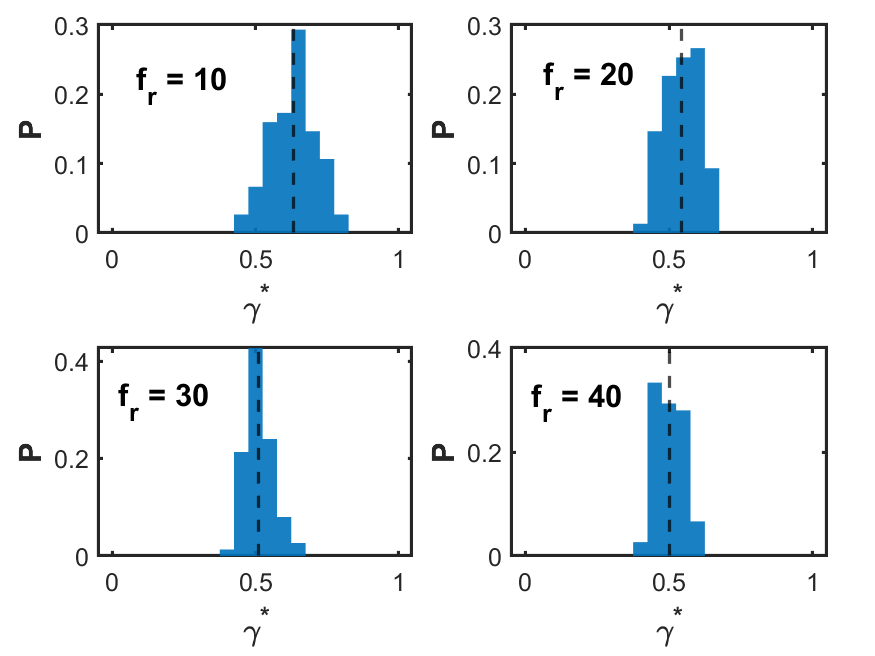}
\caption{Distribution of the optimal tournament bias \(\gamma^{*}\) for different \(f_r\), corresponding to the heat maps in Fig.~\ref{fig:gam_opt_alph_pswp}.}
\label{fig:gam_opt_distribution}
\end{figure}

\begin{equation}
\frac{\mathrm{d}}{\mathrm{d}\gamma}\,\mathbb{E}\!\left[T_r(\gamma)\right]
=\sum_{n\ge2}\Pi_n \sum_{i=1}^{n} a_i^{(n)}(\gamma)
\left(\frac{m_i^{(n)}}{\gamma}-\frac{k_i^{(n)}-m_i^{(n)}}{1-\gamma}\right),
\end{equation}
where for compactness we set
\begin{equation}
a_i^{(n)}(\gamma)=w_i^{(n)}\,F_i'\!\bigl(p_i^{(n)}(\gamma)\bigr)\,p_i^{(n)}(\gamma),
\qquad 
\end{equation}
with $F_i(p)=\sum_{c=0}^{C_0} S(f_r,p;c)\
\mathbb{P}\!\bigl(C_i^{\mathrm{eff}}=c\bigr)$. The stationary condition \(\mathrm{d}\mathbb{E}[T_r(\gamma)]/\mathrm{d}\gamma=0\) yields the fixed-point relation
\begin{equation}\label{eq:gamma_star_mixture}
\gamma
=\frac{\displaystyle\sum_{n\ge2}\Pi_n \sum_{i=1}^{n} a_i^{(n)}(\gamma)\,m_i^{(n)}}
{\displaystyle\sum_{n\ge2}\Pi_n \sum_{i=1}^{n} a_i^{(n)}(\gamma)\,k_i^{(n)}}.
\end{equation}
The factor \(a_i(\gamma)=w_i\,F_i'(\cdot)\,p_i(\gamma)\) is the \emph{marginal throughput gain} obtained by biasing selection probability toward path \(i\). Equation~\eqref{eq:gamma_star_mixture} is implicit since the marginal weights \(a_i^{(n)}(\gamma)\) depend on \(\gamma\). We therefore solve the fixed–point condition in Eq.~\eqref{eq:gamma_star_mixture} using the fitted hop profile and the empirical distribution of $n_t$ and evaluate the optimum $\gamma^{\!*}_{\mathrm{an}}(f_r)$. In parallel, the numerical optimum $\gamma^{\!*}_{\mathrm{num}}(f_r)=\arg\max_{\gamma} T_r(\gamma;f_r)$ is obtained from the simulated curves $T_r(\gamma;f_r)$. The red markers in Fig.~\ref{gam_optimal} report $\gamma^{\!*}_{\mathrm{num}}$ together with $95\%$ indistinguishable–optimal intervals computed from paired per–window differences $D_t(\gamma)=T_t(\gamma^{\!*}_{\mathrm{num}};f_r)-T_t(\gamma;f_r)$; a grid point $\gamma$ is included when $\bar D \le 1.96\,s_D/\sqrt{T}$, where $\bar D=\tfrac{1}{T}\sum_t D_t$ and $s_D^2=\tfrac{1}{T-1}\sum_t (D_t-\bar D)^2$. The blue markers display $\gamma^{\!*}_{\mathrm{an}}$ obtained from the fixed point equation. For $f_r > 10$ the two estimates agree within uncertainties and both remain concentrated near one half, with a mild drift toward larger $\gamma$ values for lighter load ($f_{r} = 10$). This is consistent with the emergence of a broad capacity plateau around $\gamma \approx \frac{1}{2}$, where $\partial_\gamma \mathbb{E}[T_r]$ is small and distributing selections across the available edge–disjoint routes is nearly optimal over a wide range of loads.

\begin{figure}[!htbp]
\centering
\includegraphics[width=0.95\columnwidth]{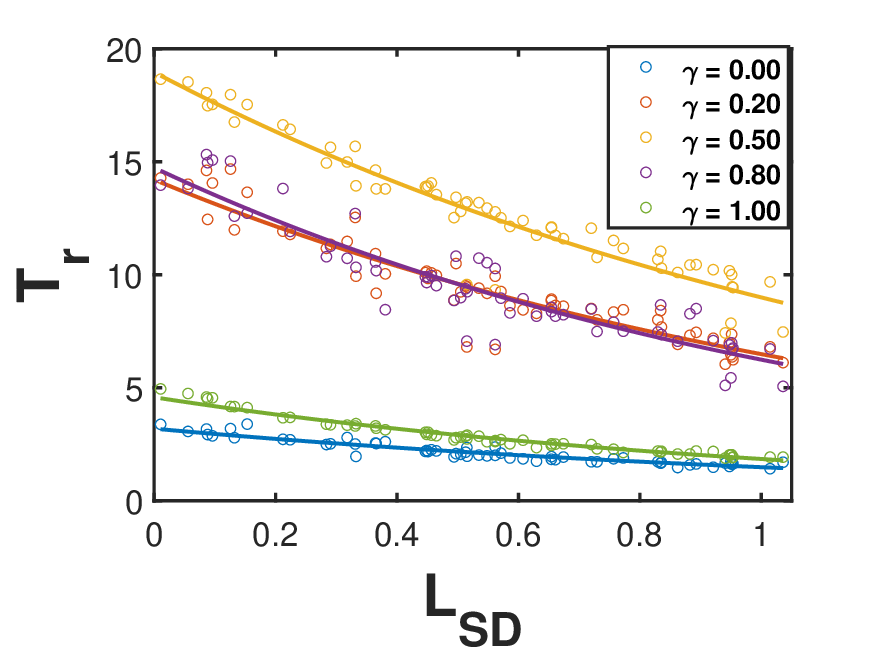}
\caption{$T_{r}$ distance scaling of the tournament routing policy for different bias values $\gamma$. Symbols show simulation data and solid lines the exponential fits $T_r(L_{SD};\gamma)\propto e^{-\kappa(\gamma)L_{SD}}$. All other parameters are identical to Fig.~\ref{Tr_sn_sim} with $f_{r} = 20$.}
\label{fig:rate_scaling}
\end{figure}

\begin{figure*}[htb]
\centering
\setlength{\tabcolsep}{1pt} 
\renewcommand{\arraystretch}{0.75} 
\begin{tabular}{cccc} 
    \includegraphics[width=0.436\textwidth]{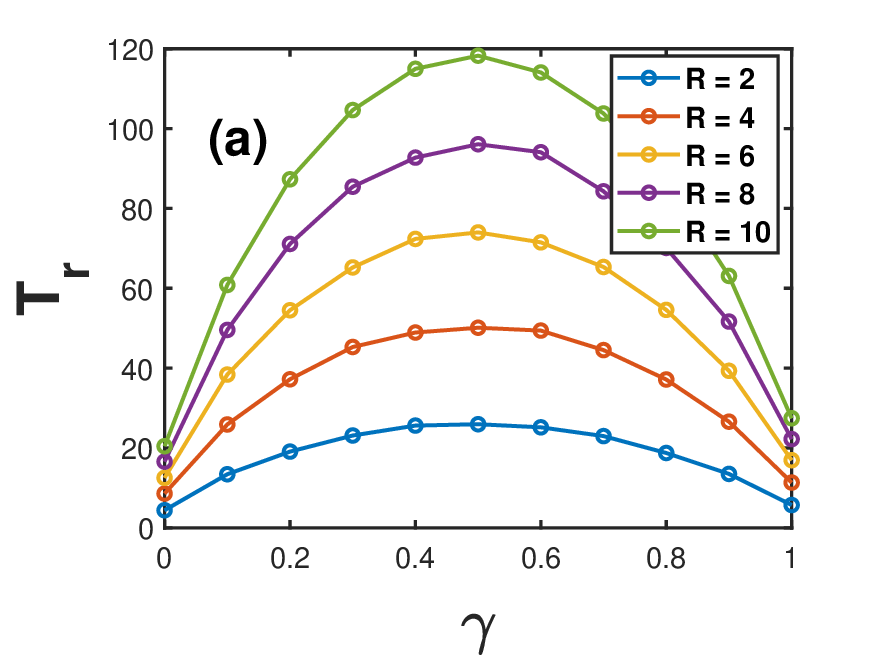} & 
    \includegraphics[width=0.436\textwidth]{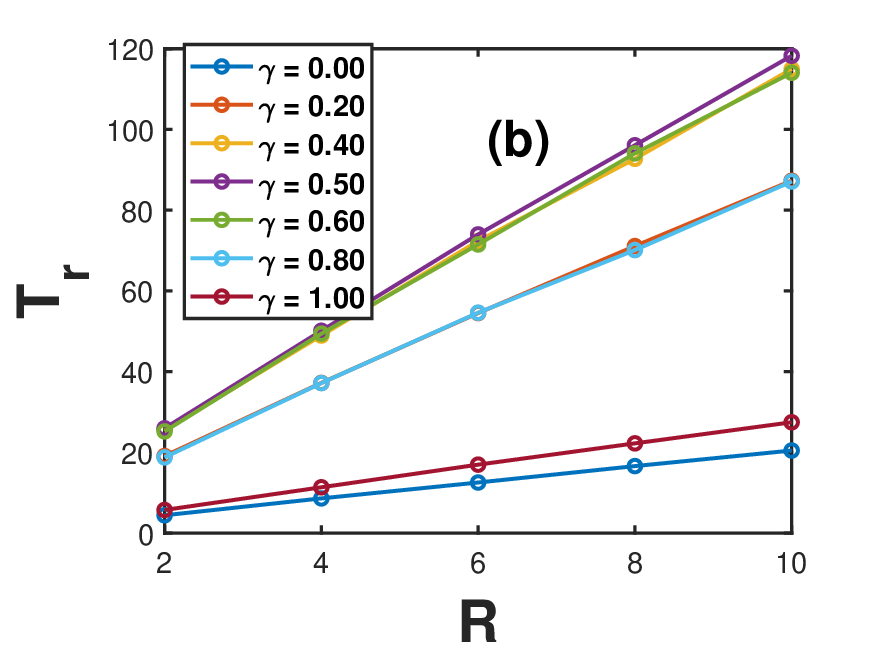} & 
\end{tabular}
\caption{Multi-request performance of tournament routing. (a) $T_r$ as a function of the tournament bias $\gamma$ for different numbers $R$ of concurrent S$-$D pairs at each time window. (b) shows $T_r$ as a function of $R$ for several fixed values of $\gamma$. Network and physical parameters are identical to those used in Fig.~\ref{Tr_sn_sim} with $f_{r} = 20$ for each pair.}
\label{multi_reqst}
\end{figure*}

Fig.~\ref{fig:gam_opt_alph_pswp} shows heat maps of the optimal tournament bias $\gamma^{\ast}$ over the $(\alpha,p_{\mathrm{swap}})$ plane for request loads $f_r \in \{10,20,30,40\}$. At light load $f_r = 10$ the optimum is shifted toward larger $\gamma$, so the tournament favors shorter paths. This is strongest when $p_{\mathrm{swap}}$ is small, since the weight $w_i = p_{\mathrm{swap}}^{h_i-1}$ strongly suppresses long paths while the shortest ones still have plenty of free capacity. As the load increases $f_{r}>10$, the optimum moves toward the balanced choice $\gamma \simeq \tfrac12$: additional requests quickly use up capacity on the upper paths, and it becomes better to spread traffic over more paths even though they carry smaller weights.
The attenuation coefficient $\alpha$ changes the heat maps only mildly. Its main effect is a slight push toward larger $\gamma$ at low load and small $p_{\mathrm{swap}}$, where losses along each hop matter most.
In Fig.~\ref{fig:gam_opt_distribution} we show the distributions of $\gamma^{\ast}$ over the same parameter grid. For $f_r = 10$ the distribution is biased toward large $\gamma$, consistent with a shortest-path preference set by swap losses when the network is lightly used. For $f_r > 10$ the distributions are much tighter and concentrated near $\gamma \simeq \tfrac12$, showing that at higher load the best choices of $\gamma$ cluster around a nearly balanced tournament. Overall, the two figures show $\gamma \approx \frac{1}{2}$ is an optimum tournament bias for a broad range of explored parameters.

We further compute the swap-weighted throughput per window $T_r(L_{SD};\gamma)$ as a function of the Euclidean source-destination separation $L_{SD}$, averaging over many windows and randomly chosen pairs, and fit the resulting curves to $T_r(L_{SD};\gamma)\simeq A(\gamma)\,e^{-\kappa(\gamma)L_{SD}}$. The results are shown in Fig.~\ref{fig:rate_scaling}. For all S$-$D separations probed, the curve corresponding to the intermediate tournament bias $\gamma\simeq 0.5$ lies strictly above those for all other values of $\gamma$, indicating that this choice yields the largest swap-weighted throughput at any fixed distance. From log-linear fits of the form $T_r(L_{SD};\gamma)\simeq A(\gamma)\,e^{-\kappa(\gamma)L_{SD}}$, we find that $\gamma\simeq 0.5$ simultaneously maximizes the prefactor $A(\gamma)$ and minimizes the effective attenuation coefficient $\kappa(\gamma)$, i.e., it achieves the slowest exponential decay of throughput with distance. In contrast, the deterministic extremes $\gamma=0$ and $\gamma=1$ exhibit larger $\kappa(\gamma)$ and smaller $A(\gamma)$, so purely diffusive or purely greedy path selection leads to inferior throughput$-$distance scaling compared to the stochastic tournament mixing implemented at intermediate bias.

We now extend our analysis from a single source$-$destination pair to the multi-request setting. Fig.~\ref{multi_reqst} reports the average swap-weighted throughput per window $T_r$ when $R$ independent source$-$destination pairs are served concurrently under tournament routing. Fig.~\ref{multi_reqst}(a) shows that, for each load level $R$, $T_r$ remains a unimodal function of the bias, with a clear maximum at an intermediate value $\gamma\simeq 0.5$, in direct analogy with the single-pair case. The peak height grows approximately linearly with $R$, reflecting the near-linear aggregation of throughput across pairs when residual capacities are managed by the tournament, whereas the deterministic extremes $\gamma=0$ and $\gamma=1$ achieve both lower maxima and stronger degradation at high $R$. Fig.~\ref{multi_reqst}(b) confirms that, for fixed $\gamma$, the total throughput increases almost linearly with the number of concurrent pairs, with the largest slope attained near $\gamma\simeq 0.5$ and substantially smaller slopes for the extremes. These results demonstrate that the Parrondo-type tournament bias that is optimal in the single-pair regime continues to deliver superior performance under multi-request load, maintaining high aggregate throughput while sharing a common, finite pool of entanglement resources.

In the main text we show that a Parrondo-style tournament mixing over all edge-disjoint paths for entanglement routing yields robust performance, derive a closed-form prediction for expected throughput, and validate it against simulations across loads and physical parameters. In the Supplementary we provide general upper bounds on throughput and structural bounds on the optimal tournament bias, clarifying when it concentrates near one half and how path-depth asymmetries shift it.\\

To conclude, we have demonstrated that a simple Parrondo–style mixing of the tournament branches over a modest set of edge–disjoint paths yields a robust interior optimum in swap–weighted throughput, outperforming either deterministic extreme. This effect persists across traffic load \(f_r\),  swap success \(p_{\mathrm{swap}}\), and attenuation \(\alpha\), with the maximizer \(\gamma^*\) concentrated near one half and becoming most pronounced at intermediate loads. In this regime, stochastic mixing distributes requests over multiple routes, delays bottleneck saturation, and converts individually weak strategies into a collectively superior policy; at very high loads the response flattens as the capacity ceiling is approached, consistent with the observed indistinguishable–optimal intervals. Beyond establishing the phenomenon, we have also provided a quantitative framework that connects per–window edge capacities to end–to–end throughput and validated it using numerical simulations on random geometric graphs. The proposed policy is lightweight; it requires no global state, only a randomized split at each tournament level, making it attractive for near–term quantum networking where resources are scarce and topology fluctuates.
Taken together, our results identify stochastic mixing as a practical mechanism to enhance entanglement distribution in realistic networks and chart a path toward scalable quantum routing.

\bibliographystyle{apsrev4-2}
\bibliography{pp_quantum.bib}

@article{wehner2018quantum,
  author = {Wehner, Stephanie and Elkouss, David and Hanson, Ronald},
  title = {Quantum internet: A vision for the road ahead},
  journal = {Science},
  volume = {362},
  number = {6412},
  pages = {eaam9288},
  year = {2018}
}

@article{pirandola2019end,
  author = {Pirandola, Stefano},
  title = {End-to-end capacities of a quantum communication network},
  journal = {Commun. Phys.},
  volume = {2},
  number = {1},
  pages = {51},
  year = {2019}
}

@article{pant2019routing,
  author = {Pant, Mohsen and Krovi, Hari and Towsley, Don and Englund, Dirk and Guha, Saikat and Tassiulas, Leandros and Jiang, Liang and Basu, Pradeep and B{\"a}uml, Stefan and others},
  title = {Routing entanglement in the quantum internet},
  journal = {npj Quantum Inf.},
  volume = {5},
  number = {1},
  pages = {25},
  year = {2019}
}

@article{bapat2023advantages,
  author = {Bapat, Aniruddha and Childs, Andrew M. and Gorshkov, Alexey V. and Schoute, Eddie},
  title = {Advantages and limitations of quantum routing},
  journal = {PRX Quantum},
  volume = {4},
  pages = {010313},
  year = {2023}
}

@article{harmer2002parrondo,
  author = {Harmer, G. P. and Abbott, D.},
  title = {Parrondo's paradox},
  journal = {Stat. Sci.},
  volume = {17},
  pages = {206--213},
  year = {2002}
}

@article{harney2025practical,
  author = {Harney, Cillian and Pirandola, Stefano},
  title = {Practical routing and criticality in large-scale quantum communication networks},
  journal = {Phys. Rev. Res.},
  volume = {7},
  number = {4},
  pages = {043168},
  year = {2025}
}

@inproceedings{shi2020concurrent,
  author = {Shi, Shouqian and Qian, Chen},
  title = {Concurrent entanglement routing for quantum networks: Model and designs},
  booktitle = {Proc. ACM SIGCOMM},
  pages = {62--75},
  year = {2020}
}

@article{azuma2023quantum,
  author = {Azuma, Koji and Economou, Sophia E. and Elkouss, David and Hilaire, Paul and Jiang, Liang and Lo, Hoi-Kwong and Tzitrin, Ilan},
  title = {Quantum repeaters: From quantum networks to the quantum internet},
  journal = {Rev. Mod. Phys.},
  volume = {95},
  number = {4},
  pages = {045006},
  year = {2023}
}

@article{pirandola2009direct,
  author = {Pirandola, Stefano and Garc{\'i}a-Patr{\'o}n, Raul and Braunstein, Samuel L. and Lloyd, Seth},
  title = {Direct and reverse secret-key capacities of a quantum channel},
  journal = {Phys. Rev. Lett.},
  volume = {102},
  number = {5},
  pages = {050503},
  year = {2009}
}

@article{takeoka2014fundamental,
  author = {Takeoka, Masahiro and Guha, Saikat and Wilde, Mark M.},
  title = {Fundamental rate-loss tradeoff for optical quantum key distribution},
  journal = {Nat. Commun.},
  volume = {5},
  number = {1},
  pages = {5235},
  year = {2014}
}

@article{pirandola2017fundamental,
  author = {Pirandola, Stefano and Laurenza, Riccardo and Ottaviani, Carlo and Banchi, Leonardo},
  title = {Fundamental limits of repeaterless quantum communications},
  journal = {Nat. Commun.},
  volume = {8},
  number = {1},
  pages = {15043},
  year = {2017}
}

@article{briegel1998quantum,
  author = {Briegel, H.-J. and D{\"u}r, Wolfgang and Cirac, Juan I. and Zoller, Peter},
  title = {Quantum repeaters: The role of imperfect local operations in quantum communication},
  journal = {Phys. Rev. Lett.},
  volume = {81},
  number = {26},
  pages = {5932--5935},
  year = {1998}
}

@article{duan2001long,
  author = {Duan, L.-M. and Lukin, Mikhail D. and Cirac, J. Ignacio and Zoller, Peter},
  title = {Long-distance quantum communication with atomic ensembles and linear optics},
  journal = {Nature},
  volume = {414},
  number = {6862},
  pages = {413--418},
  year = {2001}
}

@article{childress2005fault,
  author = {Childress, Lilian and Taylor, J. M. and S{\o}rensen, Anders S{\o}ndberg and Lukin, Mikhail D.},
  title = {Fault-tolerant quantum repeaters with minimal physical resources and implementations based on single-photon emitters},
  journal = {Phys. Rev. A},
  volume = {72},
  number = {5},
  pages = {052330},
  year = {2005}
}

@article{yuan2008experimental,
  author = {Yuan, Zhen-Sheng and Chen, Yu-Ao and Zhao, Bo and Chen, Shuai and Schmiedmayer, J{\"o}rg and Pan, Jian-Wei},
  title = {Experimental demonstration of a {BDCZ} quantum repeater node},
  journal = {Nature},
  volume = {454},
  number = {7208},
  pages = {1098--1101},
  year = {2008}
}

@article{bernien2013heralded,
  author = {Bernien, Hannes and Hensen, Bas and Pfaff, Wolfgang and Koolstra, Gerwin and Blok, Machiel S. and Robledo, Lucio and Taminiau, Tim H. and Markham, Matthew and Twitchen, Daniel J. and Childress, Lilian and others},
  title = {Heralded entanglement between solid-state qubits separated by three metres},
  journal = {Nature},
  volume = {497},
  number = {7447},
  pages = {86--90},
  year = {2013}
}

@article{chou2005measurement,
  author = {Chou, Chin-Wen and de Riedmatten, Hugues and Felinto, Daniel and Polyakov, Sergey V. and van Enk, Steven J. and Kimble, H. Jeff},
  title = {Measurement-induced entanglement for excitation stored in remote atomic ensembles},
  journal = {Nature},
  volume = {438},
  number = {7069},
  pages = {828--832},
  year = {2005}
}

@article{li2021effective,
  author = {Li, Changhao and Li, Tianyi and Liu, Yi-Xiang and Cappellaro, Paola},
  title = {Effective routing design for remote entanglement generation on quantum networks},
  journal = {npj Quantum Inf.},
  volume = {7},
  number = {1},
  pages = {10},
  year = {2021}
}

@article{zeng2023entanglement,
  author = {Zeng, Yiming and Zhang, Jiarui and Liu, Ji and Liu, Zhenhua and Yang, Yuanyuan},
  title = {Entanglement routing design over quantum networks},
  journal = {IEEE/ACM Trans. Netw.},
  volume = {32},
  number = {1},
  pages = {352--367},
  year = {2024}
}

@article{li2022fidelity,
  author = {Li, Jian and Wang, Mingjun and Xue, Kaiping and Li, Ruidong and Yu, Nenghai and Sun, Qibin and Lu, Jun},
  title = {Fidelity-guaranteed entanglement routing in quantum networks},
  journal = {IEEE Trans. Commun.},
  volume = {70},
  number = {10},
  pages = {6748--6763},
  year = {2022}
}

@article{gyongyosi2017entanglement,
  author = {Gyongyosi, Laszlo and Imre, Sandor},
  title = {Entanglement-gradient routing for quantum networks},
  journal = {Sci. Rep.},
  volume = {7},
  number = {1},
  pages = {14255},
  year = {2017}
}

@article{mishra2024parrondo,
  author = {Mishra, Ankit and Wen, Tao and Cheong, Kang Hao},
  title = {Parrondo's paradox in network communication: A routing strategy},
  journal = {Phys. Rev. Res.},
  volume = {6},
  number = {1},
  pages = {L012037},
  year = {2024}
}

@article{mishra2024efficient,
  author = {Mishra, Ankit and Wen, Tao and Cheong, Kang Hao},
  title = {Efficient traffic management in networks with limited resources: The switching routing strategy},
  journal = {Chaos Solitons Fractals},
  volume = {181},
  pages = {114658},
  year = {2024}
}

@article{cheong_1,
  author = {Cheong, Kang Hao and Koh, Jin Ming and Jones, Michael C.},
  title = {Multicellular survival as a consequence of Parrondo's paradox},
  journal = {Proc. Natl. Acad. Sci. U.S.A.},
  volume = {115},
  number = {23},
  pages = {E5258--E5259},
  year = {2018}
}

@article{cheong_2,
  author = {Koh, Jin Ming and Cheong, Kang Hao},
  title = {New doubly-anomalous Parrondo's games suggest emergent sustainability and inequality},
  journal = {Nonlinear Dyn.},
  volume = {96},
  number = {1},
  pages = {257--266},
  year = {2019}
}

@article{cheong_3,
  author = {Cheong, Kang Hao and Wen, Tao and Benler, Sean and Koh, Jin Ming and Koonin, Eugene V.},
  title = {Alternating lysis and lysogeny is a winning strategy in bacteriophages due to Parrondo's paradox},
  journal = {Proc. Natl. Acad. Sci. U.S.A.},
  volume = {119},
  number = {13},
  pages = {e2115145119},
  year = {2022}
}

\clearpage
\onecolumngrid

\renewcommand{\thesection}{S\arabic{section}}
\renewcommand{\thesubsection}{S\arabic{section}.\arabic{subsection}}
\setcounter{figure}{0}\setcounter{table}{0}\setcounter{equation}{0}
\renewcommand{\thefigure}{S\arabic{figure}}
\renewcommand{\thetable}{S\arabic{table}}
\renewcommand{\theequation}{S\arabic{equation}}

\section*{Supplementary Material: Stochastic Multipath Routing for High-Throughput Entanglement Distribution in Quantum Repeater Networks}

\section{Upper bounds on expected throughput}\label{sec:UpperBounds}

Let paths be \(i=1,\dots,n\) with hop counts \(h_i\) and swap weights \(w_i=p_{\mathrm{swap}}^{\,h_i-1}\in(0,1]\). The effective capacities \(C_i^{\mathrm{eff}}\) and per–hop attempt distributions are as defined in the main text.

\subsection{Capacity ceiling independent of \(f_r\)}
For any selection rule and any \(\gamma\),
\begin{equation}\label{eq:s_cap}
\mathbb{E}\!\bigl[T_r(\gamma)\bigr]\le\sum_{i=1}^{n} w_i\,\mathbb{E}\!\bigl[C_i^{\mathrm{eff}}\bigr].
\end{equation}
Using a geometry–averaged hop success \(p=\langle e^{-\alpha L_e}\rangle_{e\in E}\) one obtains
\begin{equation}\label{eq:s_ECeff}
\mathbb{E}\!\bigl[C_i^{\mathrm{eff}}\bigr]
=\sum_{c=1}^{C_0}\Bigl[\mathbb{P}\!\bigl(\mathrm{Bin}(C_0,p)\ge c\bigr)\Bigr]^{h_i}.
\end{equation}

\subsection{Upper bound as a function of \(f_r\)}
For any \(\gamma\),
\begin{equation}\label{eq:s_fr_min}
\mathbb{E}\!\bigl[T_r(\gamma)\bigr]\le
\min\!\Bigl\{\,w_{\max}\,f_r,\ \sum_{i=1}^{n} w_i\,\mathbb{E}\!\bigl[C_i^{\mathrm{eff}}\bigr]\Bigr\},
\qquad
w_{\max}=\max_{1\le i\le n} w_i.
\end{equation}
A \(\gamma\)–refined envelope based on \(N_i\sim\mathrm{Bin}(f_r,p_i(\gamma))\) is
\begin{equation}\label{eq:s_gamma_envelope}
\mathbb{E}\!\bigl[T_r(\gamma)\bigr]\le
\sum_{i=1}^{n} w_i\,\min\!\Bigl\{\,f_r\,p_i(\gamma),\ \mathbb{E}\!\bigl[C_i^{\mathrm{eff}}\bigr]\Bigr\}.
\end{equation}

\subsection{Mixture over a random number of disjoint paths}
If the number of available edge–disjoint paths \(n_t\) varies by window with \(\Pi_n=\mathbb{P}(n_t=n)\),
\begin{equation}\label{eq:s_mix_cap}
\mathbb{E}\!\bigl[T_r(\gamma)\bigr]\le
\sum_{n\ge 2}\Pi_n\sum_{i=1}^{n} w_i^{(n)}\,\mathbb{E}\!\bigl[C_{i}^{\mathrm{eff},(n)}\bigr],
\end{equation}
and as a function of \(f_r\),
\begin{equation}\label{eq:s_mix_fr}
\mathbb{E}\!\bigl[T_r(\gamma)\bigr]\le
\sum_{n\ge 2}\Pi_n\min\!\Bigl\{\,w_{\max}^{(n)}\,f_r,\ \sum_{i=1}^{n} w_i^{(n)}\,\mathbb{E}\!\bigl[C_{i}^{\mathrm{eff},(n)}\bigr]\Bigr\}.
\end{equation}

\section{Performance of tournament-based routing}

\begin{figure*}[t]
\centering
\setlength{\tabcolsep}{2pt}
\begin{tabular}{cc}
\includegraphics[width=0.47\textwidth]{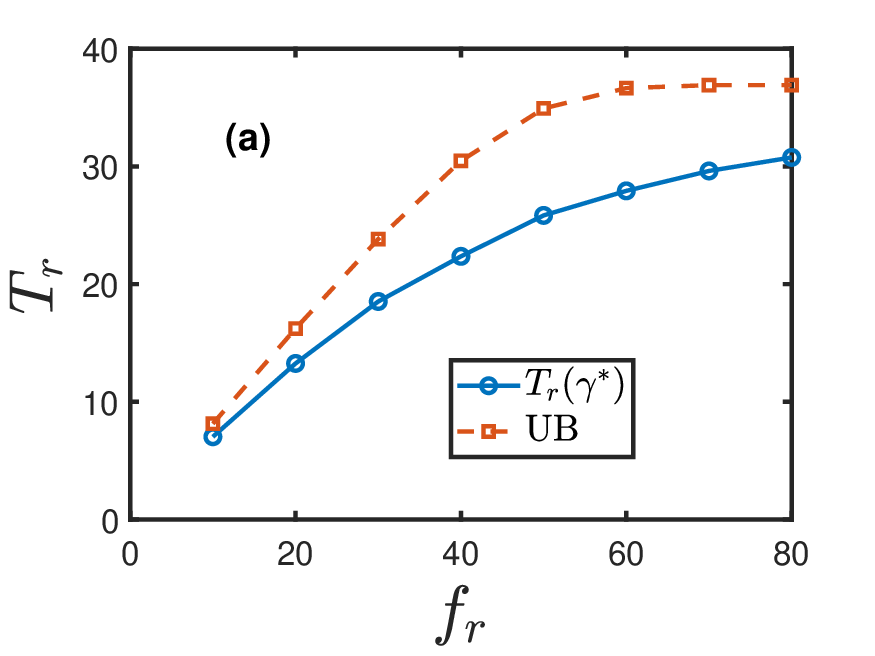} &
\includegraphics[width=0.47\textwidth]{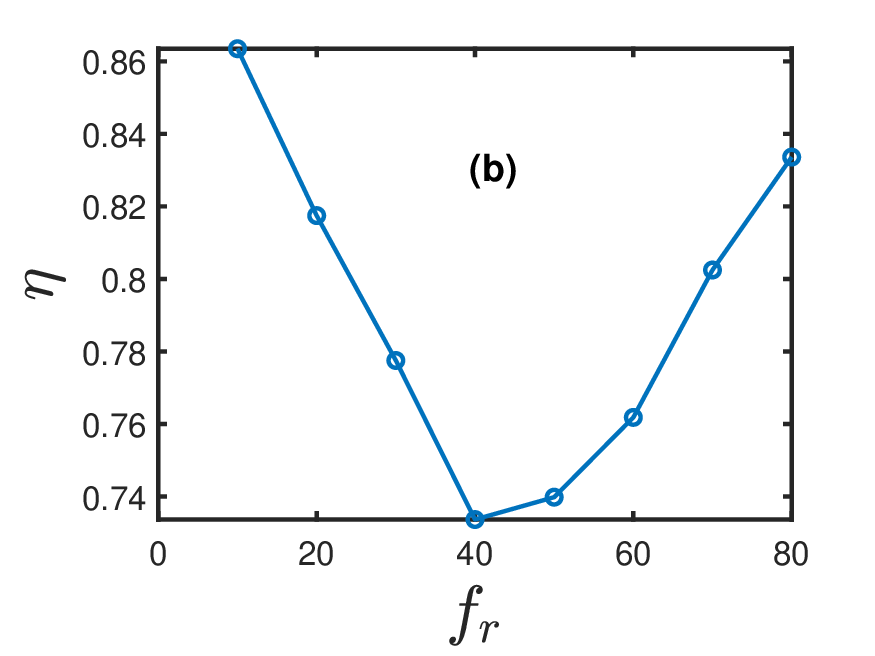}
\end{tabular}
\caption{Performance against upper bounds.  
(a) Throughput at the numerically optimal bias \(T_r(\gamma^{\!*};f_r)\) and the upper bound \(\mathrm{UB}(f_r)\) from Eq.~\eqref{eq:s_mix_fr}.  
(b) Efficiency \(\eta(f_r)=T_r(\gamma^{\!*};f_r)/\mathrm{UB}(f_r)\).  
Error bars are smaller than the markers.  Parameters match Fig.~\ref{Tr_sn_sim} of the main.}
\label{fig:UB_perf}
\end{figure*}

We benchmark the tournament policy against the capacity based envelopes of Sec.~\ref{sec:UpperBounds}. In Fig.~\ref{fig:UB_perf}(a), we plot the simulated throughput at the numerically optimal bias, \(T_r(\gamma^{\!*};f_r)\), together with the corresponding upper bound \(\mathrm{UB}(f_r)\) obtained from Eq.~\eqref{eq:s_mix_fr}. Fig.~\ref{fig:UB_perf} (b) reports the efficiency
\[
\eta(f_r)=\frac{T_r(\gamma^{\!*};f_r)}{\mathrm{UB}(f_r)}\in(0,1].
\]
For small request load the policy already captures a large fraction of the ceiling, \(\eta\simeq 0.86\).  
As \(f_r\) increases the bound grows faster than the attainable throughput because the min–term in Eq.~\eqref{eq:s_mix_fr} is dominated by the capacity side for many paths, which steepens \(\mathrm{UB}(f_r)\) and produces a mid range dip in \(\eta\).  
At higher load the curve \(T_r(\gamma^{\!*};f_r)\) approaches a capacity plateau while the bound also levels off, so \(\eta\) recovers and stabilizes, indicating that tournament mixing remains close to capacity optimal over a broad range of traffic levels.

\section{Bounds for \texorpdfstring{$\gamma^\ast$}{gamma*}}

We discuss bounds on the optimal tournament bias $\gamma^\ast$ in several representative regimes. Fix a window in which the realized number of edge–disjoint paths is $n_t=n$. From the main text the fixed–point relation reads
\begin{equation}
\gamma \;=\;
\frac{\displaystyle\sum_{i=1}^{n} a_i(\gamma)\,m_i}
     {\displaystyle\sum_{i=1}^{n} a_i(\gamma)\,k_i},
\label{eq:fixed_point_clean}
\end{equation}
where $k_i$ is the depth of leaf $i$, $m_i$ is the number of left turns along its root–to–leaf trace, and
$a_i(\gamma)=w_i\,F_i'\!\bigl(p_i(\gamma)\bigr)\,p_i(\gamma)\ge 0$ are the marginal gains defined in the main text.
It is convenient to write \eqref{eq:fixed_point_clean} as a convex combination of the per–leaf ratios $m_i/k_i\in[0,1]$. Therefore, Eqn.~\ref{eq:fixed_point_clean} immediately yields the universal enclosure
\begin{equation}
\min_{1\le i\le n} \frac{m_i}{k_i}
\;\le\; \gamma^\ast \;\le\;
\max_{1\le i\le n} \frac{m_i}{k_i}.
\label{eq:universal_bounds}
\end{equation}

\subsection{Symmetric equal–depth case}

Suppose all leaves have the same depth $k$ (hence $n=2^k$), and the hop statistics are identical across paths so that $a_i(\gamma)$ is independent of $i$. Then the fixed point reduces to the average of the ratios $\frac{m_i}{k_i}$. In a full depth–$k$ binary tree the multiplicity of leaves with $m$ left turns is $\binom{k}{m}$, therefore
\begin{equation}
\frac{1}{n}\sum_{i=1}^{n} m_i
= \frac{1}{2^{k}}\sum_{m=0}^{k} m \binom{k}{m}
= \frac{k}{2}
\quad\Longrightarrow\quad
\gamma^\ast=\frac{1}{2}.
\label{eq:balanced_half}
\end{equation}
Physically, complete symmetry (equal depths, equal swap weights, equal capacity tails) makes the tournament indifferent between left and right at every level, the optimal policy mixes uniformly across the $2^k$ disjoint routes.

\subsection{Imbalance and depth–capped bounds}

When depths and/or hop statistics differ, the fixed point \eqref{eq:fixed_point_clean} gives more influence to leaves with larger products $a_i(\gamma)\,k_i$, so the mixture tilts toward those subtrees and $\gamma^\ast$ shifts away from $1/2$. A useful consequence of \eqref{eq:universal_bounds} arises if every path uses at least one left and one right turn ($1\le m_i\le k_i-1$) and the depths are uniformly bounded by $k_{\max}$:
\begin{equation}
\frac{1}{k_{\max}} \;\le\; \gamma^\ast \;\le\; 1-\frac{1}{k_{\max}} .
\label{eq:kmax_bounds}
\end{equation}
This quantifies the maximum deviation from one half that is compatible with a given depth cap; the tighter the cap (smaller $k_{\max}$), the closer $\gamma^\ast$ must lie to $1/2$.

\subsection{Remark: relation to Kraft’s inequality}

For any rooted binary tree with leaf depths $\{k_i\}$ one has Kraft’s inequality
\begin{equation}
\sum_{i=1}^{n} 2^{-k_i} \;\le\; 1,
\label{eq:kraft}
\end{equation}
which implies $k_{\max}\ge \lceil \log_2 n\rceil$. Substituting the \emph{minimal feasible} depth cap $k_{\max}=\lceil \log_2 n\rceil$ into \eqref{eq:kmax_bounds} gives the $n$–dependent relaxation
\begin{equation}
\frac{1}{\lceil \log_2 n\rceil} \;\le\; \gamma^\ast \;\le\; 1-\frac{1}{\lceil \log_2 n\rceil},
\label{eq:n_only_relax}
\end{equation}
for the specific class of trees whose depths actually satisfy $k_i\le \lceil \log_2 n\rceil$ (e.g., uniform–depth trees). In general, however, $k_{\max}$ can be much larger than $\lceil \log_2 n\rceil$, and \eqref{eq:kraft} by itself does not yield a nontrivial $n$–only bound beyond the enclosure \eqref{eq:universal_bounds}.

\section{Jain fairness index}

\begin{figure*}[t]
\centering
\setlength{\tabcolsep}{2pt}
\begin{tabular}{cc}
\includegraphics[width=0.72\textwidth]{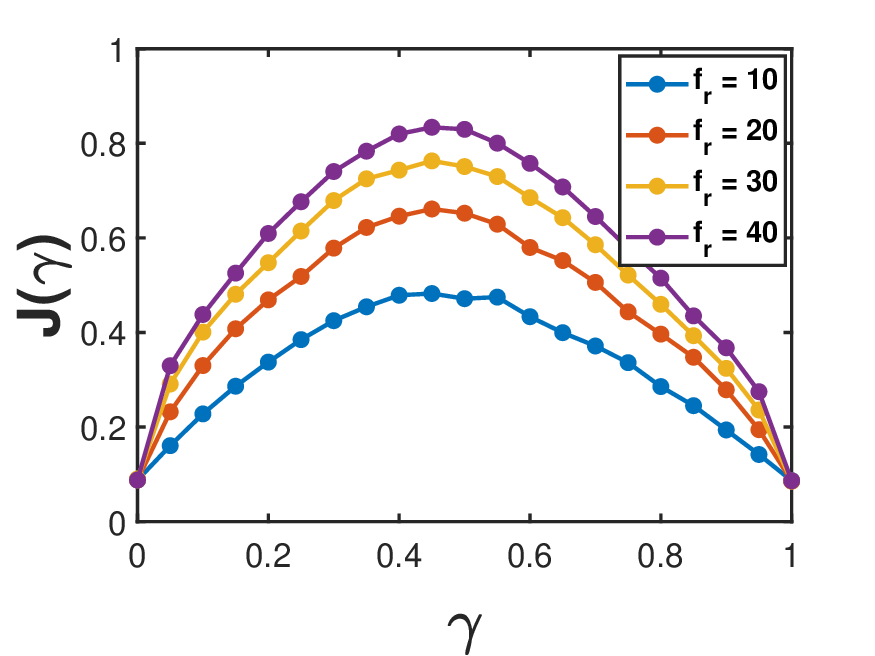} 
\end{tabular}
\caption{Jain fairness index \(J(\gamma)\) as a function of the selection bias \(\gamma\) for the tournament routing scheme. The plot quantifies how evenly the served traffic is shared across edge disjoint paths, with larger \(J(\gamma)\) indicating more balanced utilization.  The other parameters are $r=0.105$, $C_0=5$, $\alpha =1$, $T = 500$, $N=500$.}
\label{fig:jain_Delta_r}
\end{figure*}

We quantify fairness with the Jain index \(J(\gamma)\) which measures how the served traffic is shared across the edge disjoint routes. Let $u_{i,t}(\gamma)$ be the number of requests served on path $i$ in window $t$ and let $n_t$ be the number of available edge–disjoint paths in that window.
For each such window
\begin{equation}
J^{(t)}(\gamma)
=\frac{\bigl(\sum_{i=1}^{n_t} u_{i,t}(\gamma)\bigr)^{2}}{\,n_t\sum_{i=1}^{n_t} u_{i,t}(\gamma)^{2}}\,, 
\qquad
\frac{1}{n_t}\le J^{(t)}(\gamma)\le 1
\end{equation}
and the reported fairness curve is the sample mean over windows with nonzero service
\begin{equation}
{J}(\gamma)
=\frac{1}{|{T}|}\sum_{t\in{T}} J^{(t)}(\gamma)\,,
\qquad
{T}=\Bigl\{\,t:\sum_{i=1}^{n_t} u_{i,t}(\gamma)>0\,\Bigr\}.
\end{equation}
Here \(J(\gamma)=1\) indicates perfectly even sharing across all paths and \(J(\gamma)=1/n\) indicates that a single path carries the entire load.
In Fig.~\ref{fig:jain_Delta_r}, we observe a single broad maximum near \(\gamma\simeq\tfrac12\) for all \(f_r\).
For fixed \(\gamma\) the curves are ordered \(J_{40}(\gamma)>J_{30}(\gamma)>J_{20}(\gamma)>J_{10}(\gamma)\).
At \(\gamma\to0\) and \(\gamma\to1\) the index drops which signals concentration on a few routes.
Larger \(f_r\) lifts the full profile and sharpens the apex which indicates activation of deeper ranks and smoother sharing across the path family.
These trends agree with the main text near balanced selection and improved sharing at higher loads.

\section{Hop-count scaling in random geometric graphs}

\begin{figure}[!htbp]
	\centering
	\includegraphics[width= 0.72\textwidth ]{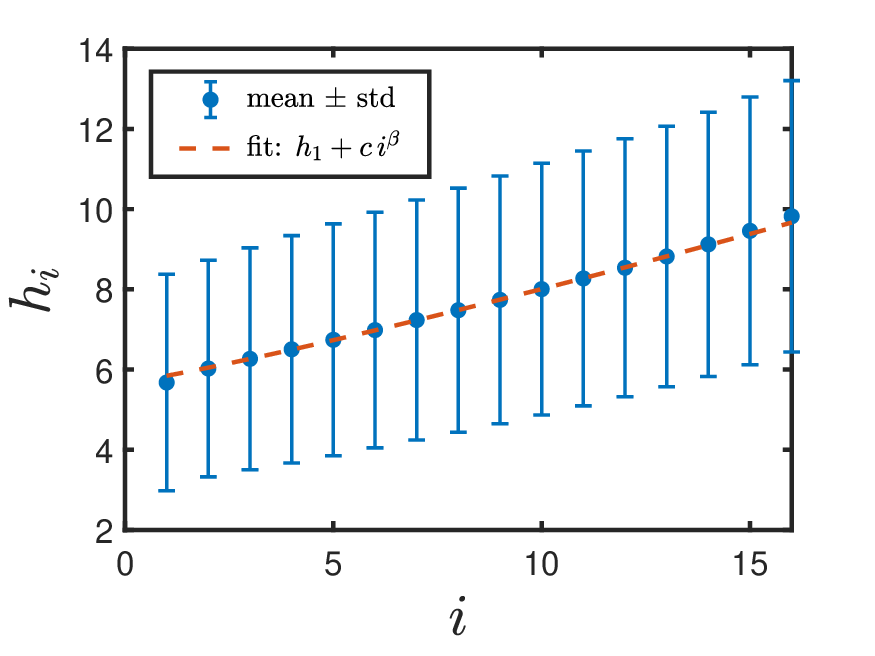} 
	\caption{Hop–count scaling in random geometric graphs. Markers show the ensemble mean \(\bar h_i\) with standard–deviation bars for ranks \(i=1,\dots,16\) (15{,}000 samples). The dashed curve is the offset power–law fit \(h_i = h_1 + c\,i^{\beta}\); $i>1$ with \(h_1=\bar h_1=5.677\), yielding \(c=0.168 [0.1518, 0.1842]\) and \(\beta=1.1422[1.1039, 1.1806]\) (95\% CIs), \(R^2=0.9984\), and RMSE \(=0.046\). }

	\label{hps_fit}
\end{figure}

Fig.~\ref{hps_fit} shows the ensemble-averaged hop counts \(h_i\) for the \(i\)-th edge-disjoint path (markers with standard-deviation bars), ranked by increasing number of hops, together with the offset power-law fit \(h_i = h_1 + c\,i^{\beta}\) ; $i>1$ (dashed).
The probability that a link carries at least one entangled pair is \(\Pr[C_e>0]=1-(1-p_{\mathrm{link}}^{(e)})^{C_0}\) with \(p_{\mathrm{link}}^{(e)}=\exp(-\alpha L_e)\). For \(\alpha=1\) we have \(\Pr[C_e>0]\approx 1\) over the observed range of \(L_e\) in the random geometric graph, so the feasible post-entanglement graph \(G_t=\{e\,|\,C_e^{(t)}>0\}\) retains the raw topology with overwhelming probability. Consequently, using hop profiles \(h_i\) fitted on the post-entanglement geometry is appropriate, and since the two topologies coincide with high probability, it is equally justified to use the pre-entanglement fit for \(h_i\) when evaluating Eq.~(\ref{eq:total_expectation_final}) of the main.

\end{document}